\documentstyle[12pt]{article}
\begin{document} 
\thispagestyle{empty} 
\begin{flushright}
UA/NPPS-4-2002\\

\end{flushright}
\begin{center}
{\large{\bf CONSERVATION OF PARTICLE MULTIPLICITIES
BETWEEN CHEMICAL AND THERMAL FREEZE-OUT\\}}
\vspace{2cm} 
{\large A. S. Kapoyannis}\\ 
\smallskip 
{\it University of Athens, Division of Nuclear and Particle Physics,\\ 
GR-15771 Athens, Greece}\\ 
\vspace{1cm}

\end{center}
\vspace{0.5cm}
\begin{abstract}
The evolution of a hadronic system after its chemical decomposition is
described through a model that conserves the hadronic multiplicities to
their values at chemical freeze-out. In the partition function describing the
model all known hadronic resonances with masses up to 2400 MeV have been
included. The state of the system is found as function of temperature and the
corresponding baryon density is evaluated. The baryon density at thermal
decoupling is also computed.
\end{abstract}

\vspace{7cm}


\newpage
\setcounter{page}{1}

{\large \bf 1. Introduction}

Thermal approaches have extensively been used to describe the particle
multiplicities which emerge from high energy collisions [1-16]. The results of
such approaches are satisfactory since they are able to predict quite
accurately a large number of different experimentally measured hadronic
multiplicities as function of a few thermodynamic variables, such as
temperature, volume and chemical potentials.

The extracted parameters from such approaches define the ``chemical
freeze-out'', i.e. the point where the chemical composition of the system
that produces the particle multiplicities is fixed. Another point called
``thermal freeze-out'' can also be defined. This second point is associated
with the particle momentum distribution which is measured experimentally.
After this point this distribution remains fixed and the particles no longer
interact among themselves.

There is evidence that in a lot of circumstances the two points corresponding
to the same system do not occur at the same temperature. Generally the
freeze-out temperature is lower than the chemical freeze-out one. Since the
particles are measured once they have reached the experimental apparatus,
after any kind of interaction among themselves has ceased and since the
thermodynamic parameters of the chemical freeze-out point predict quite well
all those particles, one has to infer that all these abundances will have
to remain fixed through the whole process between chemical and thermal
freeze-out.

In this paper the main focus will be to construct a model of relativistic
hadronic particles formulated in the grand canonical ensemble that will be
able to conserve the particle multiplicities between points which correspond
to different temperatures. Of course the two points with greatest interest are
the chemical and the freeze-out point. The question that arises then is why
there is need for another model and why not use one of the existing thermal
models. In these models the particles are described as thermally equilibrated
entities. In some of these models the hadrons are non-interacting particles
[2-7] and in others a kind of interaction among them has been included
[1,8-16]. But usually the free thermodynamic parameters are the volume $V$,
the temperature $T$ and a set of a few chemical potentials each of which is
associated with the conservation of quantum numbers like baryon number $B$,
charge $Q$, strangeness $S$, etc. During the evolution of the system among
states with different temperatures its content can alter. This is done by
adjusting the chemical potentials appropriately so as to keep the relevant
quantum numbers fixed. The only parameter that remains unfixed after the
conservation of quantum numbers is the volume. If there is need to fix the
particle numbers as well it is clear that this cannot be accomplished with
the existing parameters. Even if it is assumed that conservation of all the
particle numbers automatically conserves the quantum numbers, meaning that
only the conservation of the particles is enough, this limits the application
of the above models to only situations where the number of the particle
entities is equal to the number of the chemical potentials plus one\footnote{
In such a case the quantum number potentials and the volume would have to be
adjusted to keep the particles fixed between points with different
temperatures.}.  But generally the hadrons which have to be considered are
more numerous than the quantum numbers potentials.

As it is evident these thermal models cannot evolve the system from chemical
to thermal freeze-out. In this work the necessity to have fixed particle
numbers will be used to construct a model which will determine the evolution
of the hadronic system after its chemical freeze-out. This newly constructed
model will coincide at chemical freeze-out with ``Ideal Hadron Gas'' model
(IHG) [2-5], one of the aforementioned thermal models, which is formulated in
the grand canonical ensemble and describes hadrons as relativistic
non-interacting particles.

\vspace{0.3cm}
{\large \bf 2. Formulation of the model}

Before discussing the new model, the IHG model will be presented. In the
context of IHG the grand canonical partition function, formulated
in the Boltzmann approximation, has the form
\begin{equation}
\ln Z(V,T,\{\lambda\})_{IHG} = V \sum_{i} \lambda_{QN_i} \sum_j Z_{H_{ij}}(T)
\equiv V \sum_{i} \lambda_{QN_i} \sum_{j} \frac{T}{2\pi^2} g_{ij} m_{ij}^2
K_2(\frac{m_{ij}}{T})\;,
\end{equation}
where $i$ runs over all hadronic families such as mesons, $N$ Baryons,
$\Lambda$ Baryons, etc. and $j$ represents the specific member of the
family with degeneracy factor $g_{ij}$ and mass $m_{ij}$. $\lambda_{QN_i}$
stands for the product of all the fugacities associated with the particular
family. These fugacities can either be quantum numbers fugacities related to
Baryon number, Strangeness, etc. or to quark flavour\footnote{For example,
for $\Xi^-$ Baryons, $\lambda_{QN}$ would read
$\lambda_B\lambda_Q^{-1}\lambda_S^{-2}\gamma_s^2$ or
$\lambda_d\lambda_s^2\gamma_s^2$. One can look in [16], eq. 14,
to find out how the two sets of fugacities are related.}.

One can evaluate particle abundances if the above partition function
is extended by the introduction of a fugacity $\lambda_{ij}$ for every
particle. After calculating the particle number one has to set in IHG
$\lambda_{ij}=1$ [17], so again the particle number is only expressed as
function of the quantum numbers fugacities.

Now, if someone wishes to keep the particle numbers fixed 
it is only natural to extend (1) by the use of fugacities $\lambda_{ij}$
(corresponding to particle numbers), but with the difference that the
constraint $\lambda_{ij} = 1$ will not be imposed.
This model will be called Fixed Particle Numbers (FPN) model and
accordingly its partition function will depend on $\lambda_{ij}$'s
\begin{equation}
\ln Z(V,T,\{\lambda\})_{FPN} = V \sum_{ij} \lambda_{H_{ij}} Z_{H_{ij}}(T)
\equiv V \sum_{ij} \lambda_{H_{ij}} \frac{T}{2\pi^2} g_{ij} m_{ij}^2
K_2(\frac{m_{ij}}{T})\;,
\end{equation}
where $\lambda_{H_{ij}}$ is product of quantum numbers as well as particle
number fugacities\footnote{For example,
for $\Xi(1530)^-$ Baryon, $\lambda_{H}$ would read
$\lambda_B\lambda_Q^{-1}\lambda_S^{-2}\gamma_s^2\lambda_{\Xi(1530)^-}$.}.
Since the experimentally measured multiplicities usually contain feeding from
the decay of resonances, all known hadrons with masses up to 2400 MeV have
been included in the FPN partition function. The same hadrons have also been
included to IHG partition function (1). The mean particle number can be
evaluated through the relation
\begin{equation}
<N_{ij}>_{FPN}=\left.\lambda_{ij}
\frac{\partial \ln Z(V,T,\{\lambda\})_{FPN}}{\partial\lambda_{ij}}
\right|_{\{\lambda\}\neq\lambda_{ij}}\;,
\end{equation}
where $\{\lambda\}\neq\lambda_{ij}$ means that for the evaluation of the
partial derivative all fugacities except $\lambda_{ij}$ are considered as
constants.

The next point that has to be elucidated is at what values the particle
numbers will stay fixed. Since after chemical freeze-out these values do not
alter it is useful to formulate FPN so as to keep the particle numbers fixed
at their chemical freeze-out values. One has to remember that the chemical
freeze-out values are extracted from a thermal model, like IHG, by a
successful fit to the experimentally measured values. The particle numbers
can then be fixed at the values calculated through IHG for the chemical
freeze-out thermodynamic variables. Thus the results of IHG and FPN should
coincide at chemical freeze-out point. So it has to be
required\footnote{The primed variables in this paper will generally be
related to subsequent points of the chemical freeze-out point.}
\[
<N_{ij}>_{IHG}=<N_{ij}'>_{FPN}\Leftrightarrow
V\lambda_{QN_i}Z_{H_{ij}}(T) =
V'\lambda_{H_{ij}}Z_{H_{ij}}(T')\Leftrightarrow
\]
\begin{equation}
\lambda_{H_{ij}}=
\frac{V\lambda_{QN_i}Z_{H_{ij}}(T)}{V'Z_{H_{ij}}(T')}\;.
\end{equation}
The above equation can be used to calculate the total product of fugacities
$\lambda_{H_{ij}}$ consisting of quantum numbers fugacities and of hadron
fugacities at temperature $T'$. It has to be pointed out that it is not
possible to evaluate each quantum number fugacity separately, but this is
irrelevant since the full product of fugacities can be calculated. Another
focal point is that all quantum numbers are automatically conserved as linear
combination of the particle numbers.

In the right hand side of eq. (4) the only term which is left undetermined
after the imposition of the conservation of particle numbers is the
multiplicand factor $\frac{V}{V'}$. So an additional constraint has to be
applied. For example conservation of entropy can be assumed\footnote{A lot
of authors assume isentropic evolution of the system, e.g. see [18].}.

The entropy of the system can in general be calculated from\footnote{The
symbol of entropy is tilded in order not to be confused with the symbol of
Strangeness. $K$ can be set equal to one.}
\begin{equation}
\tilde{S}=-\left(\frac{\partial[-T\ln Z(V,T,\{\mu\})]}{\partial T}
\right)_{V,\{\mu\}}\;,
\end{equation}
where $\mu$ represents the chemical potential associated with fugacity
$\lambda=\exp(\mu/T)$. Applying (5) to the partition function
(2)\footnote{Since the IHG and FPN partition functions coincide at chemical
freeze-out point, the IHG partition function (1), where $\lambda_{ij}=1$, can
be used for the evaluation of the entropy at this point.} the
constraint of fixed entropy will read
\[
\tilde{S}=\tilde{S}'\Leftrightarrow
\]
\[
\Leftrightarrow \ln Z_{IHG}(V,T,\{\mu\})+
VT \sum_{ij} \lambda_{QN_i} \frac{\partial Z_{H_{ij}}(T)}{\partial T}-
VT \sum_{ij} \lambda_{QN_i} \frac{\mu_{QN_i}}{T^2} Z_{H_{ij}}(T)=
\]
\[
=\ln Z_{FPN}(V',T',\{\mu'\})+
V'T' \sum_{ij} \lambda_{H_{ij}} \frac{\partial Z_{H_{ij}}(T')}{\partial T'}-
V'T' \sum_{ij} \lambda_{H_{ij}}\frac{\mu_{H_{ij}}}{T'^2} Z_{H_{ij}}(T')\;.
\]
With the use of (4) the last equation becomes
\[
VT \sum_{ij} \lambda_{QN_i} \frac{\partial Z_{H_{ij}}(T)}{\partial T}-
V \sum_{ij} \lambda_{QN_i} \ln(\lambda_{QN_i}) Z_{H_{ij}}(T)=
\hspace{7cm}
\]
\[
=V'T' \sum_{ij} \frac{V\lambda_{QN_i}Z_{H_{ij}}(T)}{V'Z_{H_{ij}}(T')}
\frac{\partial Z_{H_{ij}}(T')}{\partial T'}-
V' \sum_{ij} Z_{H_{ij}}(T')
\frac{V\lambda_{QN_i}Z_{H_{ij}}(T)}{V'Z_{H_{ij}}(T')}
\ln(\frac{V\lambda_{QN_i}Z_{H_{ij}}(T)}{V'Z_{H_{ij}}(T')})\Leftrightarrow
\]
\[
\Leftrightarrow
T \sum_{ij} \lambda_{QN_i} \frac{\partial Z_{H_{ij}}(T)}{\partial T}-
\sum_{ij} \lambda_{QN_i} \ln(\lambda_{QN_i}) Z_{H_{ij}}(T) =
\hspace{7cm}
\]
\begin{equation}
\hspace{3cm}
=T' \sum_{ij} \lambda_{QN_i} \frac{Z_{H_{ij}}(T)}{Z_{H_{ij}}(T')}
\frac{\partial Z_{H_{ij}}(T')}{\partial T'}-
\sum_{ij} \lambda_{QN_i} Z_{H_{ij}}(T) 
\ln(\frac{V\lambda_{QN_i} Z_{H_{ij}}(T)}{V'Z_{H_{ij}}(T')}) \;.
\end{equation}
Setting $x\equiv\frac{V'}{V}$, (6) can be solved for $x$ to
give
\[
x_{FPN}=
\hspace{15cm}
\]
\begin{equation}
\exp\left[\frac
{\sum_{ij} \lambda_{QN_i} Z_{H_{ij}}(T) 
\ln(\frac{Z_{H_{ij}}(T)}{Z_{H_{ij}}(T')})+
T \sum_{ij} \lambda_{QN_i} \frac{\partial Z_{H_{ij}}(T)}{\partial T}-
T' \sum_{ij} \lambda_{QN_i} \frac{Z_{H_{ij}}(T)}{Z_{H_{ij}}(T')}
\frac{\partial Z_{H_{ij}}(T')}{\partial T'}}
{\sum_{ij} \lambda_{QN_i} Z_{H_{ij}}(T)}
\right].
\end{equation}

Equation (7) can be used to evaluate the volume expansion ratio as the
system has cooled to a temperature $T'$ less than the chemical freeze-out
temperature $T$. With the use of the same equation, quantities like the
baryon density of the system can be calculated at $T'$. The baryon chemical
potential at which the system is found at temperature $T'$ cannot be
calculated separately from the rest of chemical potentials in the context of
FPN. But the baryon density has no problem to be evaluated. One has to
remember that baryon number is also fixed with the imposition of the
constraints (4). So

\begin{equation}
n_{B\;FPN}=\frac{<B'>}{V'}=\frac{<B>^{ch}}{V'}=
\frac{V}{V'}\cdot\frac{<B>^{ch}}{V}=
\frac{n_B^{ch}}{x_{FPN}}\;,
\end{equation}
where $n_B^{ch}$ is the baryon density calculated at chemical freeze-out.

\vspace{0.3cm}
{\large \bf 3. Application}

The newly constructed model, FPN, can describe thermally equilibrated
hadronic systems with fixed particle numbers when their temperature is known.
As an example the systems formed at different interactions at SPS will be
considered. For this reason the chemical freeze-out parameters obtained for
these systems through fits to their experimentally measured values will be
used. These parameters are listed in Table 1 along with the references where
they can be found. From a variety of thermal analyses performed by different
authors the particular ones have been chosen because they allow for partial
strangeness equilibrium ($\gamma_s\neq1$) and they use most recent available
values for the experimentally measured hadronic multiplicities. The values of
Table 1 are then taken, for each interaction separately, as input to the
equations $<S>=0$ and $\frac{<B>}{2<Q>}=\beta$,\footnote{$\beta$ is fixed from
the baryon number and charge of the participant nucleons, e.g. see [16].}
to determine the rest of the fugacities. Thus the whole
set of chemical freeze-out parameters $(T,\mu_B,\mu_Q,\mu_S,\gamma_s)$ are
calculated and also the products of fugacities $\lambda_{QN_i}$ in (1) are
also set.

Giving different values to temperature $T$, equation (8) can be used to
calculate the corresponding baryon density. The resulting paths for FPN for
$S+S$, $S+Ag$ and $Pb+Pb$ interactions are shown in Figure 1 with solid
curves. For the $Pb+Pb$ interaction the thermal freeze-out temperature is
calculated in Refs. [20] and [21]. For these values baryon density at thermal
freeze-out $n_B^{ther}$ can be evaluated. The results are listed in the last
column of Table 2. The path for $Pb+Pb$ is followed until the lower
temperature (of the two given in Refs. [20,21]) is reached. The points which
correspond to the thermal freeze-out temperatures of these references are
depicted with squares on the FPN curve.

FPN has the unique attribute to conserve each particle species separately. So
it is not possible to compare its results directly with another thermal model.
In order to have a general view we shall depict on the graphs with the FPN
results IHG states for different temperatures. The IHG model to be used only
conserves the quantum numbers $<B>$, $<Q>$ and $<S>$ and the entropy
$<\tilde{S}>$. The IHG states are represented by a dotted curve in Figure 1.
It has to be pointed out that IHG does not take the system from chemical to
thermal freeze-out. The system remains in the context of IHG all the time at
chemical equilibrium where the particles can transform into one another.

The usefulness of Figure 1 is that, as can be observed, a ``soup'' of fixed
particles cools more efficiently than a chemically equilibrated IHG
state. This means that at the same baryon density (which through the
conservation of the baryon number is equivalent to equal volumes) smaller
temperature corresponds to the FPN state than to the IHG state. The reason is
that IHG only conserves a few quantum numbers. As temperature drops the
number of particles diminishes but without affecting the preservation of
quantum numbers. For example an equal reduction to the number of protons and
antiprotons will not affect the conservation of baryon number. In the FPN
case, on the contrary, no particle number is allowed to diminish. So in the
IHG state the available energy has to be distributed among less particles
than FPN, and their mean kinetic energy has to be greater, leading to greater
temperature.

In Figure 2 the FPN ratios $x=V'/V$, where $V$ is the chemical freeze-out
volume, are plotted as function of temperature for the three SPS
interactions. For comparison the ratios $x=V'_{IHG}/V_{IHG}$ for the
particular IHG model discussed above are plotted with dotted curves. The two
volumes in the IHG ratios correspond to chemically equilibrated states at
different temperatures. $V'_{IHG}$ and $V_{IHG}$ are calculated for the same
temperatures as $V'$ and $V$ respectively.

Also, in order to show the effect of hadrons with large masses, FPN
calculations with hadrons only up to the Delta mass have been included in
figures 1 and 2. The chemical freeze-out parameters for $Pb+Pb$ of Table 1
have been used for these calculations. From figure 1 it is evident that the
baryon density calculated through the truncated version of FPN is considerably
less than the baryon density calculated at the same temperature with the FPN
model using all the hadrons. The expansion ratio $x$ for the truncated FPN,
though, is close to the calculated ratio with FPN using all the hadrons, as
it can be seen from figure 2.

\vspace{0.3cm}
{\large \bf 4. Conclusion}

After chemical freeze-out the collisions among hadrons that compose the
hadronic gas can no longer change its chemical composition. Following this
requirement a non-interacting hadron gas model (FPN) has been presented that keeps the
multiplicity of every particle fixed to the value dictated by the chemical
freeze-out conditions. In the context of FPN the constraints of conservation
of quantum numbers are broken up to a larger number of constraints, these of
conservation of particle numbers. The chemical potentials of quantum numbers
are no longer ``good'' variables to describe the evolution of the system. Of
course the fugacities of particle numbers used as variables in FPN are not
``free'' parameters. Their values are fixed from the given set of the quantum
numbers fugacities at chemical freeze-out. So the evolution of a hadronic
system is described as function of temperature (after imposing conservation
of entropy). This is done for three SPS interactions.

Following this evolution and using values of thermal freeze-out temperature
extracted for the $Pb+Pb$ interaction the baryon density at freeze-out is
evaluated. As the temperature at thermal decoupling for various interactions
can be calculated using transverse mass spectra or HBT analysis [22] the same
procedure can be applied to evaluate the corresponding baryon density before
free streaming for these interactions.

In this paper the IHG has been used as the thermal model which would coincide
with FPN at chemical freeze-out point. Any other thermal model, interacting
or non-interacting, can also be used in place of IHG and, with the use of
particle fugacities, a model that conserves particle multiplicities can also
be formed. Recently in model II of [23]\footnote{[23] appeared during the
refereeing period of the present paper.} the particle fugacities have been
used to conserve particle multiplicities. This a three dimensional
hydrodynamic model describing radial and elliptic flow but it includes
hadrons with masses only up to Delta mass and it is restricted to zero baryon
chemical potential.

\vspace{0.3cm}
{\large \bf Acknowledgement}
I would like to thank Professor N. G. Antoniou and Professor C. N. Ktorides
for reviewing the manuscript and for useful remarks.

\newpage
{\large{\bf Figure Captions}}
\newtheorem{g}{Figure} 
\begin{g}
\rm Contours (solid lines) that follow hadronic systems after chemical
freeze-out on ($T,n_B$) plane for 3 interactions at SPS, calculated through
FPN (model of Fixed Particle Numbers).
On the same graph points at chemical equilibrium (dotted lines) calculated
through an IHG model that conserves baryon number, charge, strangeness and
entropy are depicted.
The slashed curve represents calculations with FPN including hadrons with
masses only up to the Delta mass (FPN (a)) and using as chemical freeze-out
parameters the ones that correspond to $Pb+Pb$ interaction of Table 1.

\end{g}
\begin{g}
\rm The ratios of the volumes $V'$ of the SPS hadronic systems at a certain
temperature to their volumes $V^{ch}$ at chemical freeze-out calculated for
FPN (solid lines). The dotted lines correspond to the ratios of volumes
of chemically equilibrated states of IHG model of Figure 1 calculated at the
same temperatures as the volumes of the FPN ratios. The slashed curve
represents calculations with model FPN (a) of Figure 1.
\end{g}

\newpage
\vspace{0.3cm}
{\large{\bf Table Captions}} 
\newtheorem{f}{Table} 
\begin{f} 
\rm Chemical freeze-out parameters calculated for different interactions
at SPS and the corresponding references.
\end{f}
\begin{f} 
\rm Thermal freeze-out temperature calculated in two different references
for the $Pb+Pb$ interaction and the corresponding computation of baryon
density through FPN. The upper errors of baryon density correspond to the
upper errors of temperature. The same is true for the lower errors.
\end{f}

\vspace{0.3cm}
\begin{center}
\begin{tabular}{|c|ccc|c|} \hline
Experiment & $T^{ch} (MeV)$ & $\mu_B^{ch} (MeV) $ & $\gamma_s^{ch}$ &
 Reference \\ \hline\hline
S+S 200 $A\cdot GeV$ & $180.5\pm10.9$ & $220.2\pm18.0$ & $0.747\pm0.048$ & [7,18]\\
S+Ag 200 $A\cdot GeV$ & $178.9\pm8.1$ & $241.5\pm14.5$ & $0.711\pm0.063$ & [7,18]\\
Pb+Pb 158 $A\cdot GeV$ & $174.7\pm6.7$ & $240\pm14$    & $0.900\pm0.049$ & [19]\\
\hline
\end{tabular} 
\end{center} 

\begin{center}
Table 1.
\end {center} 

\vspace{0.3cm}
\begin{center}
\begin{tabular}{|c|cc|c|} \hline
Experiment & $T^{ther} (MeV)$ & Reference& $n_B^{ther}$ ($fm^{-3}$) \\
\hline\hline
Pb+Pb 158 $A\cdot GeV$ & $120\pm12$   & [20] & $0.099^{+0.022}_{-0.019}$ \\
Pb+Pb 158 $A\cdot GeV$ & $95.8\pm3.5$ & [21] & $0.0627^{+0.0047}_{-0.0045}$ \\
\hline
\end{tabular} 
\end{center} 

\begin{center}
Table 2.
\end {center}


\begin{thebibliography}{99}
\bibitem{1} J. Cleymans, H. Satz, Z. Phys. C 57 (1993) 135.
\bibitem{2} J. Cleymans, K. Redlich, H. Satz, E. Suhonen,
Z. Phys. C 58 (1993) 347.
\bibitem{3} J. Sollfrank, M. Ga\'{z}dzicki, U. Heinz, J. Rafelski, Z. 
Phys. C 61 (1994) 659.
\bibitem{4} J. Letessier, A. Tounsi, U. Heinz, J. Sollfrank,
J. Rafelski, Phys. Rev. D 51 (1995) 3408.
\bibitem{5} A. D. Panagiotou, G. Mavromanolakis, J. Tzoulis, Phys.
Rev. C 53 (1996) 1353.
\bibitem{6} F. Becattini, U. Heinz, Z. Phys. C 76 (1997) 269.
\bibitem{7} F. Becattini, M. Ga\'{z}dzicki, J. Sollfrank, Eur. Phys. J.
C 5 (1998) 143.
\bibitem{8} D. H. Rischke, M. I. Gorenstein, H. St$\ddot{\rm o}$cker,
W. Greiner, Z. Phys. C 51 (1991) 485.
\bibitem{9} R. Hagedorn, Nuovo Cimento Suppl. III (1965) 147.
\bibitem{10} R. Hagedorn, J. Ranft, Nuovo Cimento Suppl. VI (1968) 169;
R. Hagedorn, Nuovo Cimento Suppl. VI (1968) 311.
\bibitem{11} R. Hagedorn, Nuovo Cimento LVI A (1968) 1027.
\bibitem{12} R. Hagedorn, J. Rafelski, Phys. Lett. B 97 (1980) 136.
\bibitem{13} A. S. Kapoyannis, C. N. Ktorides, A. D. Panagiotou,
J. Phys. G 23 (1997) 1921.
\bibitem{14} A. S. Kapoyannis, C. N. Ktorides, A. D. Panagiotou, 
Phys. Rev. D 58 (1998) 034009.
\bibitem{15} A. S. Kapoyannis, C. N. Ktorides, A. D. Panagiotou, 
Phys. Rev. C 58 (1998) 2879.
\bibitem{16} A. S. Kapoyannis, C. N. Ktorides, A. D. Panagiotou, 
Eur. Phys. J. C 14 (2000) 299.
\bibitem{17} R. Hagedorn, K. Redlich, Z. Phys. C 27 (1985) 541.
\bibitem{18} J. Cleymans, K. Redlich, Phys. Rev. C 60 (1999) 054908.
\bibitem{19} F. Becattini, J. Cleymans, A. Ker$\ddot{\rm a}$nen, E.Suhonen,
K. Redlich, Phys. Rev. C 64 (2001) 024901.
\bibitem{20} H. Appelsh$\ddot{\rm a}$user et al., NA49 Collaboration,
Nucl. Phys. A 638 (1998) 91c; Eur. Phys. J. C 2 (1998) 661.
\bibitem{21} J. R. Nix et al., nucl-th/9801045.
\bibitem{22} U. Heinz, J. Phys. G 25 (1999) 263; R. Stock, Nucl. Phys.
A 661 (1999) 282c.
\bibitem{23} T. Hirano, K. Tsuda, talk given at 30th International Workshop
on Gross Properties of Nuclei and Nuclear Excitation: Hirschegg 2002,
in Hirschegg 2002, Ultrarelativistic heavy-ion collisions 152-157,
e-Print Archive: nucl-th/0202033.
\end{thebibliography}
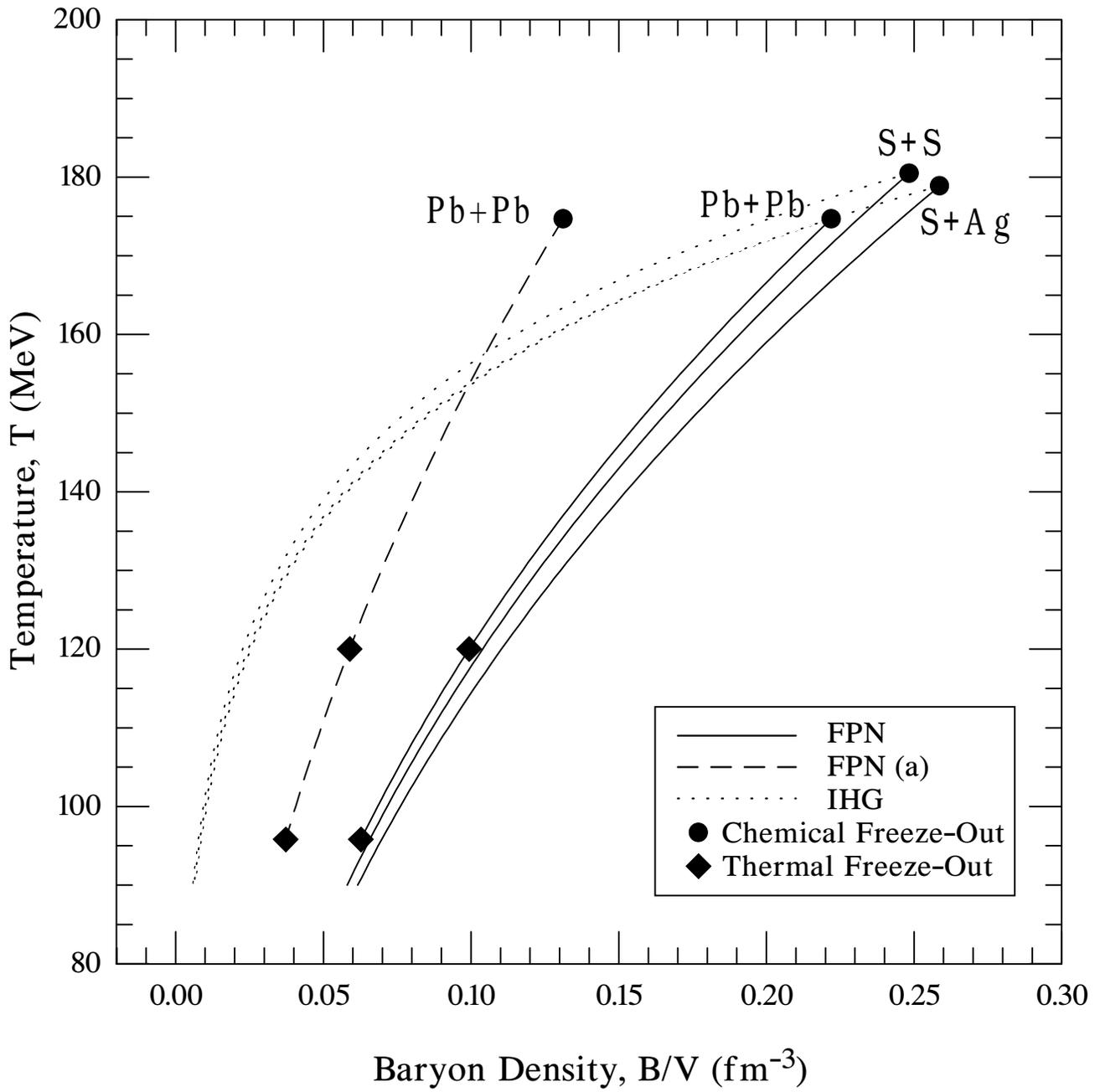
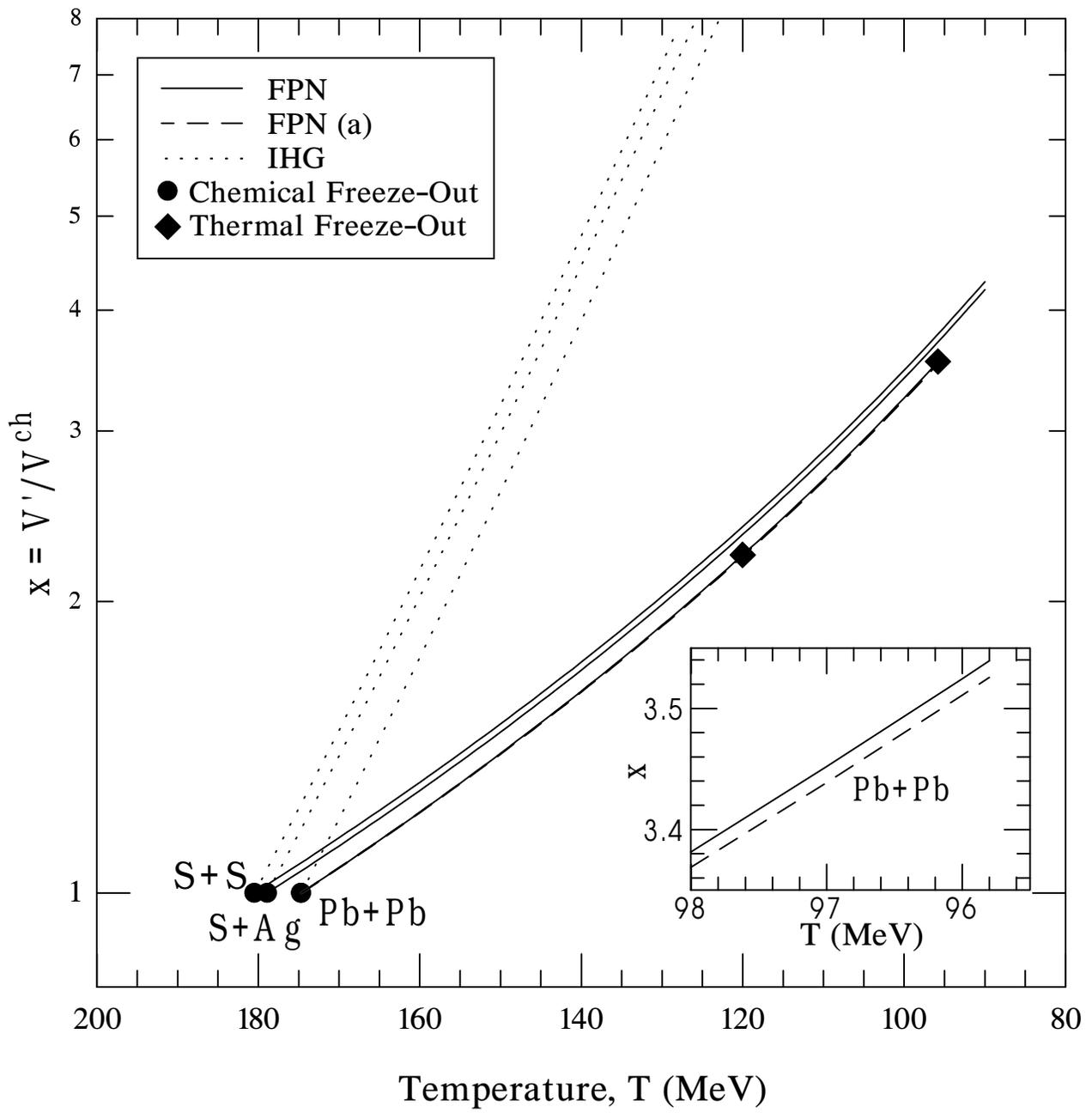
\end{document}